# Error correction and fast detectors implemented by ultra-fast neuronal plasticity

Roni Vardi,[1*+] Hagar Marmari[1*] and Ido Kanter[1,2+]

[1]Gonda Interdisciplinary Brain Research Center and the Goodman Faculty of Life Sciences, Bar-Ilan University, Ramat-Gan 52900, Israel

[2]Department of Physics, Bar-Ilan University, Ramat-Gan 52900, Israel

We experimentally show that the neuron functions as a precise time-integrator, where the accumulated changes in neuronal response latencies, under complex and random stimulation patterns, are solely a function of a global quantity, the average time-lag between stimulations. In contrast, momentary leaps in the neuronal response latency follow trends of consecutive stimulations, indicating ultra-fast neuronal plasticity. On a circuit level, this ultra-fast neuronal plasticity phenomenon implements error-correction mechanisms and fast detectors for misplaced stimulations. Additionally, at moderate/high stimulation rates this phenomenon destabilizes/stabilizes a periodic neuronal activity disrupted by misplaced stimulations.

## I. INTRODUCTION

On the network and circuit level, both synaptic and neuronal plasticity are present. These two distinct types of plasticity have different effects on the dynamics of a network [1, 2]. On one hand, synaptic plasticity has been vastly researched, from the single neuron to the network level, specifically long and short-term plasticity. Short-term synaptic plasticity reflects an increase (facilitation) and decrease (depression) in the probability of neurotransmitter release [3, 4]. It affects the speed of synaptic signal transmission and can last from hundreds of milliseconds to seconds [3, 5]. This phenomenon varies enormously depending on the neuronal and synaptic features as well as on the neuron's recent history of activity [6-8]. Neuronal plasticity, on the other hand, was examined mainly on the single neuron level [9, 10], hence investigation of this phenomenon is still demanded.

Short-term synaptic plasticity, in the form of facilitation and depression (FAD), is suggested to carry critical computational functions in neural circuits [1, 11, 12], thus one can hypothesize that neuronal plasticity carries similar computational functions. This hypothesis was not experimentally verified on the network level, and in addition its enormous variation seems to prevent reliable information processing [13, 14]. Here we experimentally demonstrate neuronal ultra-fast plasticity on a time scale of several milliseconds and its applications to advanced computational tasks.

## II. FAD ON A SINGLE NEURON LEVEL

At the single neuron level, one of the most significant time-dependent features is the neuronal response latency, L, to ongoing stimulations, which is measured by the time-lag from the beginning of a stimulation to its corresponding evoked spike [15, 16]. When a neuron is stimulated repeatedly at a frequency typically exceeding ~1 Hz, its response latency stretches gradually. The accumulated stretching over few hundreds of repeated stimulations is typically several milliseconds, increasing with the stimulation rate, and terminating at the intermittent phase, where the latency fluctuates around an average value [15-18]. This slow latency increase is a fully reversible phenomenon, which decays substantially after a waiting time of few seconds without stimulations, and in a timescale of several minutes the initial response latency is restored.

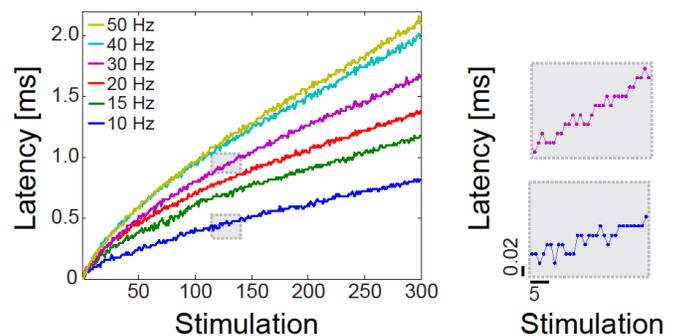

FIG. 1 (color online). Experimental measurements of the response latency of a single neuron at 10, 15, 20, 30, 40 and 50 Hz (increasing frequency from bottom to top line). Fluctuations in the neuronal response latency at 10 and 30 Hz are exemplified by the zoom-in (gray areas, bottom and top respectively).

To exemplify this neuronal feature, stimulations were given to cultured cortical neurons that were functionally isolated from their network by pharmacological blockers of both excitatory and inhibitory synapses [15, 17]. Stimulations at a fixed frequency in the range [10, 50] Hz indicate an increase of ~0.8-2.1 ms, relative to the initial neuronal response latency, over the course of 300 stimulations, representing a form of depression (Fig. 1). Eventually the neuron reaches a constant average latency value, at the intermittent phase (not shown) similar to [16, 17]. Although the neuronal response latency increases on the average, locally it can increase, decrease or remain unchanged under 20 μs resolution (Table 1 and gray areas of Fig. 1). The overall increase in the neuronal response latency, depression, is attributed to the following two factors. The probability for a local increase in the neuronal response latency is slightly greater than the probability for a local decrease (Table 1). In addition, the average increase in the neuronal response latency per evoked spike is slightly greater than the average decrease (Table 1). The bias in these two factors is enhanced as the stimulation frequency is increased, and accordingly the depression amplitude increases.

| Frequency [Hz] | Total latency increase [ms] | Probability of a local decrease in latency | Probability of a local increase in latency | Mean latency decrease per evoked spike [μs] | Mean latency increase per evoked spike [μs] |
|---|---|---|---|---|---|
| 10 | 0.82 | 0.43 | 0.57 | -14.52 | 15.64 |
| 15 | 1.18 | 0.41 | 0.59 | -14.40 | 16.79 |
| 20 | 1.36 | 0.40 | 0.60 | -14.16 | 17.14 |
| 30 | 1.66 | 0.42 | 0.58 | -14.80 | 20.39 |
| 40 | 1.98 | 0.38 | 0.62 | -18.18 | 21.90 |
| 50 | 2.10 | 0.38 | 0.62 | -16.14 | 21.70 |

Table 1. Statistical analysis of the local changes in neuronal response latency of the neuron presented in Fig. 1, over a course of 300 stimulations. The neuron was periodically stimulated at rates in the range of [10, 50] Hz, and its response latencies were calculated using voltage minima estimation (see Section VI). The global change in the neuronal response latency over the course of 300 stimulations is defined as ΔL= L(300)-L(1) and the local change in the neuronal response latency is defined as ΔL(i)= L(i)-L(i-1), where L(i) stands for the neuronal response latency at the i$^{th}$ stimulation. The table shows the probability for a local increase/decrease in the neuronal response latency, calculated as P(ΔL(i)>0) / P(ΔL(i)<0), and the mean increase/decrease in latency per evoked spike, calculated as mean(ΔL(i)>0) / mean(ΔL(i)<0) μs. It appears that with the increase in stimulation frequency, P(ΔL(i)>0))-P(ΔL(i)<0) increases, as well as mean(ΔL(i)>0) + mean(ΔL(i)<0). These two trends are responsible for the rising amplitude of the overall depression with stimulation rate.

The underlying mechanism to quantitatively measure the trends of FAD on a single neuron level is the unavoidable changes in the neuronal response latency to ongoing stimulations. For a momentary increase/decrease in the stimulation frequency, facilitation/depression is observed through the latency of the neuronal response to the misplaced stimulations (Fig. 2a). FAD is measured by the momentary latency leap corresponding to the sudden change in the stimulation frequency (Fig. 2a). Typically the amplitude of FAD, increases with the momentary frequency change, however, at high stimulation rates the effect of facilitation diminishes and even might vanish (Fig. 2b).

The amplitude of FAD significantly varies with the timing of the misplaced stimulation and among neurons. Nevertheless, we find a robust and systematic global feature governing depression for a given neuron under a complex stimulation pattern. The profile of the neuronal response latency under a complex stimulation pattern follows the profile of the neuronal response latency under a fixed stimulation rate, with the same average time-lag between stimulations (Fig. 2c). For instance, the neuronal response latency under alternating stimulations at 10 Hz (100 ms) and 30 Hz (~33.3 ms) follows the response latency profile under a fixed stimulation rate where the time-lag between stimulations is ~66.7 ms, equivalent to a stimulation rate of 15 Hz, which differs from the average frequency, 20 Hz (Fig. 2c). This systematic global feature of a neuron functioning as a precise time-integrator was found to be applicable even for random stimulation patterns (Fig. 2d).

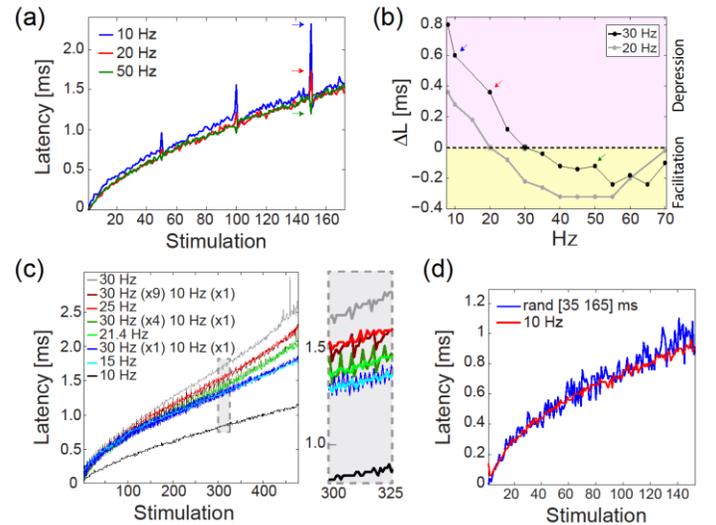

FIG. 2 (color online). (a) Experimental measurements of momentary changes in the neuronal response latency under stimulation at 30 Hz, where stimulations 50, 100, 150 were given at 10, 20 and 50 Hz (indicated by upper, middle and lower arrow, respectively). (b) A single neuron was stimulated at a frequency of 20 Hz (gray (lower) line) and 30 Hz (black (upper) line), where at stimulation number 150 a different frequency in the range of [8, 70] Hz was given, resulting in a latency leap $\Delta L = L(150) - L(149)$. Arrows correspond to latency leaps indicated in (a). (c) The neuronal response latency under stimulation at 10 Hz (black (lower) line), 30 Hz (gray (upper) line) and at 30 Hz where the frequency changed to 10 Hz every m=2 (dark blue (lower) jittered line), 5 (dark green (middle) jittered line), 10 (dark red (upper) jittered line) stimulations, and respectively at periodic stimulations 15 (light blue (lower) smooth line), 21.4 (light green (middle) smooth line) and 25 (light red (upper) smooth line) Hz. The right panel shows a zoom-in of the gray area. (d) The neuronal response latency under stimulation at 10 Hz (red smooth line) and at random time-lags between stimulations in the range of [35, 165] ms (blue jittered line).

## III. FAD UTILIZES COMPUTATIONAL FUNCTIONS

The local change in the neuronal response latency, $\Delta L(i) = L(i) - L(i-1)$ (where $L(i)$ stands for the neuronal response latency at the $i^{th}$ stimulation), can increase (depression), decrease (facilitation) or remain unchanged under 20 μs time resolution (Fig. 1 and Table 1). The probability for a consistency between local trends in FAD ($\Delta L$) and the difference between two consecutive time-lags ($\Delta \tau$) is measured by $P_+ = P(\Delta L \cdot \Delta \tau > 0)$. It can reach ~0.85 for wide ranges of random time-lags between stimulations (Fig. 3a-b), and decreases for narrower ranges (e.g. ~0.6 for N1 in Fig. 3d and upper panel of Fig. 3e). Hence, FAD responds momentarily with high probability to local trends of a random stimulation pattern. Note that the consistency, $P_+$, was verified to be very similar between trials, however the precise timings of the inconsistencies varied.

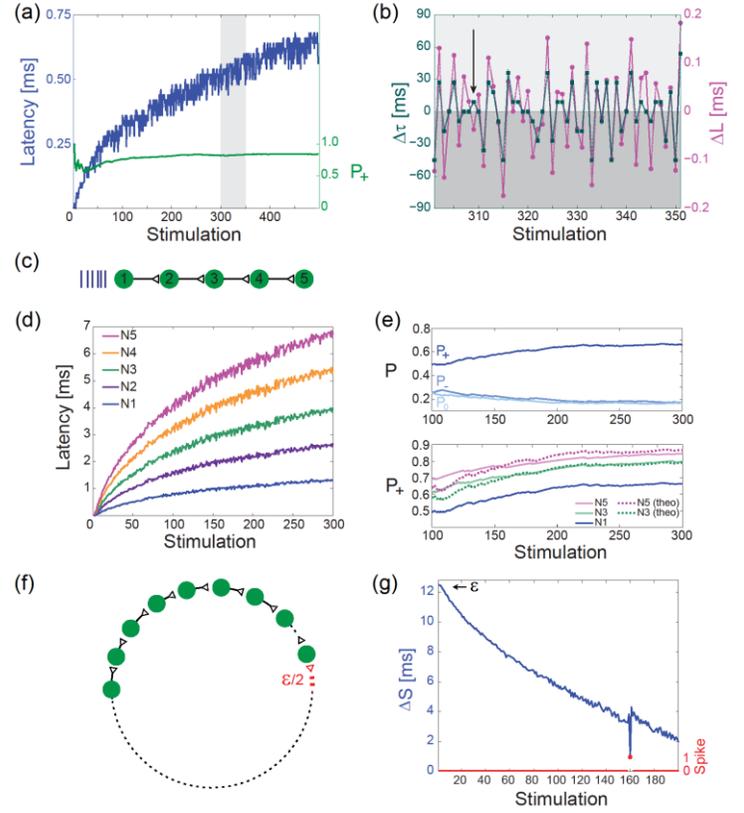

FIG. 3 (color online). (a) The neuronal response latency stimulated with random time-lags in [50, 150] ms (blue) and the accumulated probability $P_+ = P(\Delta L \cdot \Delta \tau > 0)$ (purple). (b) A zoom-in (gray area in (a)) of $\Delta L$ (pink circles) and the corresponding $\Delta \tau$ (aqua squares). For most steps $\Delta L \cdot \Delta \tau > 0$, with an exception at step 309 (black arrow), in this trial. (c) Schematic of a chain consisting of five neurons, where neuron 1 is stimulated by a random stimulation pattern. (d) The accumulated response latency for the first Ni neurons along the chain, where the lines are ordered from bottom to top from i=1 to i=5. (e) Top panel: accumulated probabilities $P_+$, $P_-$ and $P_0$ for N1 in (d). Bottom panel: accumulated $P_+$ for N1 (blue (lower) full line), N3 (green (middle) full line) and N5 (purple (upper) full line) in (d), and the theoretically predicted $P_+$ for N3 (purple (lower) dashed line) and N5 (green (upper) dashed line). (f) Schematic of a neuronal circuit consisting of nine neurons and strong/weak (above/sub-threshold) stimulations represented by full/dashed lines. The rightmost neuron receives two weak stimulations via two delay routes with an initial difference of ε ms (red wide dashed line). (g) Experimental results for an initial ε=12.5 ms. The leftmost neuron is stimulated at random time-lags in the range [27.3, 39.3] ms with the exception of 100 ms preceeding stimulation 160. The time-lag between two stimulations arriving at the rightmost neuron, $\Delta S$ (blue), results in an evoked spike solely for stimulation 160 (red circle).

The hypothesis that FAD supports a variety of fast neural computations requires a reliable mechanism [19, 20]. We propose a prototypical error-correcting FAD mechanism (amplifier) based on a neuronal chain, which was experimentally confirmed using a 5-neuron chain stimulated with a random pattern (Fig. 3c-e). Since FAD of each one of the five neurons follows the local trends of the stimulation pattern with probability $P_+\sim0.6$, on a chain level the depression is enhanced, resulting in $P_+\sim0.85$ (lower panel of Fig. 3e). From the experimental measurements of the probabilities $P_+$, $P_-=P(\Delta L\cdot\Delta\tau<0)$ and $P_0=P(\Delta L\cdot\Delta\tau=0)$ one can theoretically estimate the level of error-correction of a chain using the assumptions of independent momentary FAD between neurons and a fixed positive/negative change in the neuronal response latency per spike (Fig. 1). Under these assumptions for a 3-neuron chain one can verify

$$P_+(3)=P_+^3+3P_+^2(P_0+P_-)+3P_+P_0^2$$

and for a 5-neuron chain

$$P_+(5)=P_+^5+5P_+^4(P_0+P_-)+10P_+^3(P_0+P)^2+10P_+^2(P_0^3+3P_-P_0^2)+5P_+P_0^4$$

equations which are approximated, but in a good agreement with the experimental results (Fig. 3e).

Using the enhanced FAD along a neuronal chain, we propose a neuronal circuit functioning as an instantaneous detector for a misplaced stimulation where the output neuron is stimulated by two sub-threshold stimulations (Fig. 3f and Section VI), e.g. only two temporally close stimulations result in an evoked spike, similar to an AND gate [17]. Initially, the difference between the two input-output delay routes prevents an evoked spike of the output neuron. However, a single postponed stimulation or equivalently several missing stimulations, results in an enhanced depression and an alert in the form of a single evoked spike (Fig. 3g).

The two computational building blocks experimentally demonstrated here rely on the FAD phenomenon. Similar functionalities were theoretically proposed using similar connection schemes affected by different background properties (i.e. a noisy environment, see [21] for a review). However, the robustness of both these mechanisms to a more biologically realistic framework is required. Specifically, extensive spatial summation, inhibitory synapses, and an integration of the FAD amplitudes into the model have not been addressed. Extensions to more complex models with more than one presynaptic neuron at each node and population dynamics are also required in order to provide the full extent of the computational functions.

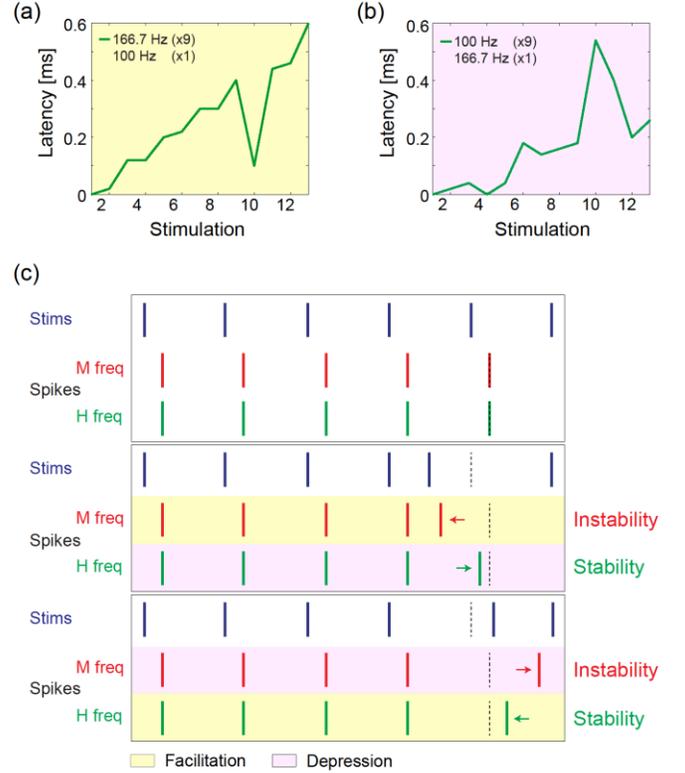

FIG. 4 (color online). (a) Experimental measurements of the response latency at 6 ms (166.7 Hz) with the exception of 10 ms (100 Hz) preceding the 10[th] stimulation. (b) A reversed scenario to (a), 10 ms (100 Hz) with the exception of 6 ms (166.7 Hz) preceding the 10[th] stimulation. (c) *Top box:* Stimulations at a fixed rate (blue), and their corresponding evoked spikes for moderate (M, red) and high (H, green) stimulation frequencies. Evoked spikes are shown as a function of stimulation number. *Middle box:* Similar to the top box, however the 5[th] stimulation is given at a *higher* rate, earlier than expected under a fixed stimulation rate (dashed vertical line). *Bottom box:* Similar to the top box, however the 5[th] stimulation is given at a *lower* rate, later than expected.

## IV. FAD MAINTAINS STABILITY/ INSTABILITY

For high stimulation frequencies, typically exceeding 100 Hz, the local trends of FAD are reversed. Namely, a momentary decrease in the stimulation frequency results in facilitation (Fig. 4a), as opposed to depression observed under moderate stimulation frequencies (Fig. 2a and 2b). Similarly, a momentary increase in the stimulation frequency, much beyond 100 Hz, results in depression (Fig. 4b), as opposed to facilitation observed under moderate stimulation frequencies (Fig. 2a and 2b). This reversed effect at high frequencies is evident even after a very short sequence of periodic stimulations where the momentary ΔL is comparable with the entire accumulated latency increase (Fig. 4a and 4b). The appearance of depression at high stimulation frequencies (Fig. 4b) is consistent with the disappearance of facilitation at high-moderate frequencies (Fig. 2b) since FAD is expected to be continuous over the entire frequency range. In addition, this depression is much enhanced as the time-lag corresponding to the misplaced stimulation approaches the neuronal response latency of the last periodic stimulation (not shown).

Facilitation and depression are typically considered reversed mechanisms [3], hence each one is anticipated to realize different computational tasks [11]. Nevertheless, we show that FAD realizes a unified computational task. For moderate frequencies FAD leads to *instability*, where a misplaced stimulation shifts its corresponding evoked spike further from its expected timing given a periodic stimulation pattern (Fig. 4c.). For high stimulation frequencies the situation is reversed and FAD leads to *stability*, where a misplaced stimulation shifts its corresponding evoked spike towards the expected timing given a periodic stimulation pattern (Fig. 4c). This high frequency stability mechanism can be amplified by the accumulated effect of a neuronal chain, similar to the proposed error-correcting mechanism exemplified for moderate stimulation frequencies (Fig. 3c-e). In addition, the emergence of facilitation for a delayed stimulation (last row of Fig. 4c) leads to repulsion between evoked spikes resulting from two nearby stimulations; otherwise the later evoked spike eventually might be annihilated by the refractory period. Thus, this repulsion might also be attributed to a mechanism preserving the neuronal information embedded in nearby spikes.

## V. CONCLUSION

Synaptic plasticity and neuronal plasticity have different effects on a network level [1, 2]. Synaptic plasticity affects the transmission of a signal through a link, a synapse, connecting two nodes, neurons, in the network. In contrast, neuronal plasticity affects the internal dynamics of a node, neuron, in the network. Consequently, synaptic plasticity affects all transmission of information passing through a link, whereas neuronal plasticity affects all information routes passing through a node. These two phenomena have different impacts on the information flow in a network, especially when a node functions as a hub with many connections, as in the case of scale-free networks, characterized by several hubs with very high connectivity [22]. Hence, it is clear that synaptic plasticity and neuronal plasticity may have different computational implications on normal and abnormal brain functionality.

Finally, we note that in order for the proposed computational mechanisms to be applicable to brain functionalities, further investigation is required. Specifically, the robustness of the mechanism to population dynamics and in vivo recordings as well as the steady state of the neuronal response latency as opposed to the transient period shown here.

## VI. METHODS

*Culture preparation*. Cortical neurons were obtained from newborn rats (Sprague-Dawley) within 48 h after birth using mechanical and enzymatic procedures [15, 18, 23]. All procedures were in accordance with the National Institutes of Health Guide for the Care and Use of Laboratory Animals and Bar-Ilan University Guidelines for the Use and Care of Laboratory Animals in Research and were approved and supervised by the Institutional Animal Care and Use Committee. The cortical tissue was digested enzymatically with 0.05% trypsin solution in phosphate-buffered saline (Dulbecco's PBS) free of calcium and magnesium, and supplemented with 20 mM glucose, at 37°C. Enzyme treatment was terminated using heat-inactivated horse serum, and cells were then mechanically dissociated. The neurons were plated directly onto substrate-integrated multi-electrode arrays (MEAs) and allowed to develop functionally and structurally mature networks over a time period of 2-3 weeks in vitro, prior to the experiments. Variability in the number of cultured days in this range had no effect on the observed results. The number of plated neurons in a typical network was in the order of 1,300,000, covering an area of about 380 mm$^2$. The preparations were bathed in minimal essential medium (MEM-Earle, Earle's Salt Base without L-Glutamine) supplemented with heat-inactivated horse serum (5%), glutamine (0.5 mM), glucose (20 mM), and gentamicin (10 g/ml), and maintained in an atmosphere of 37°C, 5% $CO_2$ and 95% air in an incubator as well as during the electrophysiological measurements.

*Synaptic blockers.* All experiments were conducted on cultured cortical neurons that were functionally isolated from their network by a pharmacological block of glutamatergic and GABAergic synapses. For each culture 20 µl of a cocktail of synaptic blockers was used, consisting of 10 µM CNQX (6-cyano-7-nitroquinoxaline-2,3-dione), 80 µM APV (amino-5-phosphonovaleric acid) and 5 µM Bicuculline. This cocktail did not block the spontaneous network activity completely, but rather made it sparse. At least one hour was allowed for stabilization of the effect. Variability in the amount of synaptic blockers in the range of [12, 40] µl had no effect on the observed results.

*Stimulation and recording.* An array of 60 Ti/Au/TiN extracellular electrodes, 30 µm in diameter, and spaced either 200 or 500 µm from each other (Multi-ChannelSystems, Reutlingen, Germany) were used. The insulation layer (silicon nitride) was pre-treated with polyethyleneimine (0.01% in 0.1 M Borate buffer solution). A commercial setup (MEA2100-2x60-headstage, MEA2100-interface board, MCS, Reutlingen, Germany) for recording and analyzing data from two 60-electrode MEAs was used, with integrated data acquisition from 120 MEA electrodes and 8 additional analog channels, integrated filter amplifier and 3-channel current or voltage stimulus generator (for each 60 electrode array). Mono-phasic square voltage pulses ([-900, -100] mV, [40, 1500] µs) were applied through extracellular electrodes. Each channel was sampled at a frequency of 50k samples/s, thus the changes in the neuronal response latency were measured at a resolution of 20 µs.

*Cell selection.* Each node was represented by a stimulation source (source electrode) and a target for the stimulation – the recording electrode (target electrode). These electrodes (source and target) were selected as the ones that evoked well-isolated, well-formed spikes and reliable response with a high signal-to-noise ratio. This examination was done with a stimulus intensity of -800 mV using 30 repetitions at a rate of 5 Hz followed by 1200 repetitions at a rate of 10 Hz.

*Stimulation control.* A node response was defined as a spike occurring within a typical time window of 2-10 ms following the electrical stimulation [24]. The activity of the source and target electrodes of each node in the feed-forward neuronal circuit/chain was collected. Conditioned stimulations were enforced on the circuit/chain neurons, embedded within a large-scale network of cortical cells in vitro, following the circuit/chain connectivity. The timings of conditioned stimulations to the subsequent node in the feed-forward neuronal circuit/chain were computed off-line according to the timings of evoked spikes of the former node. Strong stimulations, resulting in a reliable neural response, were given in the range of ([-800, -700] mV, [40, 200] µs). Weak (sub-threshold) stimulations, given to the rightmost neuron in Fig. 3f, varied among neurons and trials and were given in the range of ([-800, -300] mV, [40, 1500] µs), such that an evoked spike is expected only if the time-lag between two consecutive weak stimulations is short enough. In the presented results, the stimulation parameters were (-800 mV, 200 µs) (strong) and (-500 mV, 1100 µs) (weak). In cases where there was a partial overlap between two consecutive weak stimulations, a single complex stimulation (imitating the structure of partially overlapped stimulations) was applied in order to overcome technical limitations.

Results were confirmed in experiments where the circuit/chain nodes were represented by different neurons as well as by the same neuron (different or the same source and target electrodes).

No correlation was found between neuronal response latencies of different neurons as well as within the same neuron, under random stimulation patterns taken from a uniform distribution. Lack of correlation was confirmed using the following steps. For a given node, the quantity $P_+^k = (\sum_{i=1} \theta(\Delta L_i^k \cdot \Delta \tau_i))/N$ was measured, where θ stands for a heaviside step function, N stands for the length of the time series stimulations and k for the measurement number. Each of the trials was recorded after a few minutes of relaxation. Next, we found that for the first and the second measurements, for instance, $P_+^1 \cdot P_+^2 \sim (\sum_{i=1} \theta(\Delta L_i^1 \cdot \Delta \tau_i) \cdot \theta(\Delta L_i^2 \cdot \Delta \tau_i))/N$. This agreement was confirmed for different ranges of random stimulation time-lags, $\{\Delta \tau_i\}$. The lack of correlations is essential for the effectiveness of error-correcting mechanisms (Fig. 3e) and is also confirmed by the fairly good agreement between the simplified theoretical estimation (see equations in the text) and the experimental results (lower panel of Fig. 3e).

*Data analysis.* Analyses were performed in a Matlab environment (MathWorks, Natwick, MA, USA). Action potentials were detected off-line by threshold crossing, voltage estimation at threshold, and voltage minima estimation. In the context of this study, no significant difference was observed in the profile of the neuronal response latency under either method of spike detection. The reported results were confirmed based on at least eight experiments each, using different sets of neurons and several tissue cultures.

*Spike detection and response latency calculation by threshold crossing*. Recordings from selected electrodes were analyzed off-line [15, 23, 25]. Spikes were detected only when the absolute value of the sampled signal passed a certain threshold level, which varied between neurons and thus determined per neuron. The neuronal response latency was calculated as the duration from the beginning of a stimulation to the first sampled point crossing the set threshold. All presented data was analyzes using this method, except for the data in Table S1.

*Spike detection and response latency calculation by estimating voltage minima with interpolation*. Recordings from selected electrodes were analyzed off-line, using a detection window of typically 2-10 ms following the beginning of a stimulation. In order to surpass the 20 μs timescale of the recording device, the following interpolation method was used. We fit a parabola to the local voltage minima ($v_2$) and the two nearby voltage recordings ($v_1$ and $v_3$). The three coefficients (a, b, c) of the interpolated parabola

$$v=at^2+bt+c$$

are determined using the three points ($t_1$,$v_1$), ($t_2$,$v_2$), ($t_3$,$v_3$) where $t_1=t_2-20$ μs and $t_3=t_2+20$ μs. One can verify that

$$t_{min}=0.5(-4v_2+3v_1+v_3)/(v_3-2v_2+v_1)$$

The neuronal response latency was then calculated as the duration from the beginning of a stimulation to $t_{min}$.
The data presented in Table 1 was analyzed using this method.

## ACKNOWLEDGMENTS
We thank Moshe Abeles for stimulating discussions. Invaluable computational assistance by Yair Sahar, Amir Goldental and technical assistance by Hana Arnon are acknowledged. This research was supported by the Ministry of Science and Technology, Israel.

* These authors equally contributed to this work
+ ido.kanter@biu.ac.il, ronivardi@gmail.com